\documentclass[aps,10pt,eqsecnum, preprint,nofootinbib,superscriptaddress]{revtex4}
\usepackage{amssymb,amsmath,amsthm,graphicx,amscd}
\usepackage{enumerate,color,changepage,ulem,bm}
\usepackage[hidelinks]{hyperref}

\def\be{\begin{equation}}
\def\te{\end{equation}}
\def\ee{\end{equation}}

\def\bea{\begin{eqnarray}}
\def\tea{\end{eqnarray}}
\def\eea{\end{eqnarray}}

\begin{document} 
 
\title{\Large Zeroth Law in Quantum Thermodynamics at Strong Coupling: `in Equilibrium', not `Equal Temperature'} 

\author{Jen-Tsung Hsiang}
\email{cosmology@gmail.com}
\affiliation{Center for High Energy and High Field Physics, National Central University, Chungli 32001, Taiwan, ROC}
\author{Bei-Lok Hu}
\email{blhu@umd.edu}
\affiliation{Maryland Center for Fundamental Physics and Joint Quantum Institute, University of Maryland, College Park, Maryland 20742-4111, USA}

\begin{abstract}
The  zeroth law of thermodynamics involves a transitivity relation (pairwise between three objects) expressed either in terms of `equal temperature' (ET), or `in equilibrium' (EQ) conditions. In conventional thermodynamics conditional on vanishingly weak system-bath coupling these two conditions are commonly regarded as equivalent. In this work we show that for thermodynamics at strong coupling they are inequivalent: namely, two systems  can be in equilibrium and yet have different effective temperatures. A recent result \cite{NEqFE} for Gaussian quantum systems shows that an effective temperature $T^{*}$ can be defined  at all times during a system's nonequilibrium evolution, but because of the inclusion of interaction energy, after equilibration the system's $T^*$ is slightly higher than the bath temperature $T_{\textsc{b}}$, with the deviation depending on the coupling. A second object coupled with a different strength with an identical bath at temperature $T_{\textsc{b}}$ will not have the same equilibrated temperature as the first object. Thus $ET \neq EQ $ for strong coupling thermodynamics. We then investigate the conditions for dynamical equilibration for two objects 1 and 2 strongly coupled with a common bath $B$, each with a different equilibrated effective temperature. We show this is possible, and prove the existence of a generalized fluctuation-dissipation relation under this configuration. This affirms that `in equilibrium' is a valid and perhaps more fundamental notion which the zeroth law for quantum thermodynamics at strong coupling should be based on. Only when the system-bath coupling becomes vanishingly weak that `temperature' appearing in thermodynamic relations becomes universally defined and makes better physical sense.   
\end{abstract}
\maketitle

\hypersetup{linktoc=all}

\newtheorem{theorem}{Theorem}
\baselineskip=18pt
\allowdisplaybreaks

\section{Introduction}

The zeroth, first and second laws of thermodynamics are usually tied to the definitions of temperature, energy and entropy, respectively \cite{Frohlich}. In this paper we are interested in the conditions for the zeroth law. What we know as the  zeroth law of thermodynamics \cite{FowGug} involves a transitivity relation~\cite{Berg} (pairwise between three objects $A$, $B$, $C$) expressed either in terms of `equal temperature' (ET), or `in equilibrium' (EQ) conditions. We shall use `ET' to denote a relation between two objects read as, `has the same temperature as' and `EQ' to denote `in equilibrium with'. The `equal temperature'  version of the zeroth law, namely, ``if $A$ ET $B$ and if $B$ ET $C$, then $A$ ET $C$" was used by Maxwell \cite{Maxwell} and  Sommerfeld~\cite{Somm}. The corresponding `in equilibrium' or EQ version appeared in Planck\footnote{The relevant passage in the 6th Edition (1921)  translated to English by Turner can be found in \cite{Turner1}.}\cite{Planck}, Carath\'eodory\footnote{For a comprehensive description of his work on thermodynamics see the review of Redlich \cite{Redlich}.}\cite{Cara} and Born \cite{Born}. 

%was first stated by Maxwell \cite{Maxwell} in 1872 (by the account of  \cite{Evans}): At equilibrium, Bodies whose temperatures are equal to that of the same body have themselves equal temperatures. Clearly at least these two elements enter in the zeroth law, one is the concept of temperature, the other is the equilibrium condition.

\subsection{The zeroth law: Temperature vs Equilibrium}

\paragraph{Temperature defined under weak coupling with bath}

The concept of temperature associated with a system in thermal equilibrium is as old as thermodynamics itself. There are many versions describing this relation. E.g., the relation described by Carath\'eodory was made rigorous by Miller \cite{Miller}, who stated it as a necessary and sufficient condition for the existence of temperature from the thermodynamic laws. As we know the temperature of a physical object can be defined in many ways in conventional thermodynamics. In microcanonical ensembles it is related to the internal energy of a system, in canonical ensembles it is related to how the number of accessible states (measured by entropy) varies with energy. Viewing temperature from  the perspective of energy and entropy  some authors find the zeroth law redundant\footnote{E.g.,  Buchdall \cite{Buch}   shows that the  transitivity relation in the zeroth law,-- i.e., ``is in diathermic equilibrium with" -- can be deduced from the first law and the second law if the latter makes no reference to temperature. Kammerlander and Renner~\cite{KamRen}, laying out the foundations of thermodynamics  starting from the   basic concepts of systems, processes and states, show that the zeroth law can be derived from the first and the second laws.}. For strong coupling thermodynamics this remains to be seen\footnote{We will  be in a better position to comment on this claim  after we examine the internal energy \cite{HMF} and the entropy \cite{QTD2} of (Gaussian) quantum systems.}. What we can say now is that already at this very basic level of the zeroth law there is ambiguity in the definition of a universal or absolute temperature -- `universal' or `absolute' (we don't mean Kelvin temperature) in the sense that it is independent of the specific features of the object,  and this situation will pervade in all strong-coupling thermodynamic laws invoking temperature. 

An important underlying assumption of conventional thermodynamics is that the system is very weakly coupled with the bath so that the system's internal energy does not contain the interaction energy. 
%The notion of an absolute temperature  is based on this assumption. In conventional (ultra-weak coupling) thermodynamics  
Under this assumption when a system is in contact with a thermal bath of temperature $T_{\textsc{b}}=\beta_{\textsc{b}}^{-1}$, under certain thermodynamic limits and barring the exceptional cases (such as systems with negative heat capacity or baths with logarithmic correlations mentioned below) it will evolve in time to a thermal state of the same temperature. These assumptions underlie the notion of a universal temperature scale, independent of the details of the specific system involved.  We simply say that the system has a certain temperature $\beta_{\textsc{b}}^{-1}$. It is universal and thus the relation is transitive in the sense that whatever system comes into  contact with the same thermal bath will reach the same temperature. However, when a system is strongly coupled to a thermal bath, the notion of a universal, absolute temperature scale is lost.  It has been known~\cite{GWT84,HTL11a,TH20} that if we prepare the combination $C=S+B$ of a system $S$, made up of a quantum Brownian oscillator, and a  quantum thermal bath $B$, with $C$ initially in a global equilibrium thermal state, then the (reduced) density matrix operator of the oscillator in general does not assume the canonical form when the system-bath coupling is finite, but not vanishingly small. This thwarts the attempt to identify a {\it bona fide} temperature for the system.

When a system is strongly coupled to a thermal bath, the temperature of this system is no longer a well-defined quantity -- it depends on how much one includes their interaction energy in the system versus the bath. Novel quantities like the Hamiltonian of mean force are introduced for this purpose. This is one of the reasons why we want to revisit all the strong coupling thermodynamic laws, starting with the zeroth law. For strong coupling at least two issues stand out: a) Can a temperature, or an effective temperature $T^{*}$ be defined? b) Suppose such an effective temperature can be defined, will the transitive relation  of isothermality (equal temperature) still hold -- namely, let object 1 at effective temperature $T_1^*$ be in equilibrium with a bath $B$ and a different object 2 at temperature $T_2^*$ be in equilibrium with an identical bath $B$, is it true that $T_1^*=T_2^*$, as is the case in conventional (weak coupling) thermodynamics?  From our recent work on Gaussian systems the answer to a) is yes: One can define an effective temperature for a system in a nonequilibrium setting. We shall show in this paper that the answer to b) is no: $T_1^* \neq T_2^* \neq T_{\textsc{b}}$  even if each object is in equilibrium with an identical bath at temperature $T_{\textsc{b}}$.  The criterion for the zeroth law in strong coupling thermodynamics should be based on the equilibrium condition, not equal temperature. 
%The statement is, if oscillator 1 is in equilibrium with a bath $B$ at $T_{\textsc{b}}$ and 2 is in equilibrium with $B$ bath at $T_{\textsc{b}}$), then 1 is in equilibrium with 2, even though the temperatures of 1 and 2 at equilibrium with the bath are not equal, that is,  $T^{*}_1 \neq T^{*}_2$. The ET version of the zeroth law becomes valid only when the coupling strength is vanishingly small. 

\paragraph{The zeroth law formulated under equilibrium conditions}

It is well known that there is no zeroth law for systems which are not, or cannot be, in equilibrium.   
Examples are systems with long range interactions \cite{Camp} or systems with negative heat capacity,  such as  $2D$ Coulomb gas, gravitating systems \cite{LBW,Mario}, polymers, glassy systems,  etc. They will be excluded from consideration here, namely, we assume ``the absence of appreciable long-range forces between different portions of all bodies" \cite{Turner2} under study.  

We note that almost all of the discussions of classical thermodynamics theories in  these earlier works are in the context of a system very weakly coupled to a thermal bath, and taking a relatively short time to relax to equilibrium. Strong coupling and baths with long correlation times can affect the equilibration time of the system significantly. For example, Patra and Bhattacharya~\cite{PatBha} studied numerically the system of one harmonic oscillator placed in different types of baths. In computationally achievable time, at low to moderate coupling strengths, their results show  a violation of the zeroth law of thermodynamics in the scenarios involving logarithmic thermostat. Specifically, ``(i) the kinetic and configurational temperatures of the systems are different, (ii) momentum distribution of log thermostat is non-Gaussian, and (iii) a temperature gradient is created between the kinetic and configurational variables of the log thermostat."

\subsection{Zeroth law for strongly coupled quantum thermodynamics}

We have recently studied the nonequilibrium (NEq) thermodynamics of a quantum Brownian harmonic oscillator, our system, which is strongly coupled to a quantum scalar field bath with bilinear coupling \cite{NEqFE}. Since the complete system is Gaussian we can solve this problem exactly and from it establish a theory of quantum thermodynamics at strong coupling, at least for Gaussian systems. From the NEq free energy of the system  based on a coarse-grained effective action and from the von Neumann entropy constructed from the reduced density matrix, we find that  a \textit{nonequilibrium effective temperature $T^*$} can indeed be defined for the quantum system at every moment of time throughout the entire evolution.  In this paper we shall use this quantum NEq systematics to address some foundational issues of the zeroth law in strong coupling quantum thermodynamics\footnote{Evans et al \cite{Evans} have also approached the zeroth law from a NEq viewpoint. In providing a proof of the zeroth law for classical, deterministic, $T$-mixing systems, they derived an exact expression for the far-from-equilibrium thermal conductivity of the material concerned.}.   

%\begin{description} 	\item[a. 
We first state the \textbf{main claims from the present work} and then describe how we are led to them.   The notion of temperature is in general not well defined for a system strongly coupled to a thermal bath~\cite{GWT84,HTL11a,MU19}. However, an effective temperature can be defined at least for strongly coupled Gaussian quantum systems, in fact, throughout its nonequilibrium evolution \cite{NEqFE}.  We will show in this paper that  \textit{two Gaussian quantum systems with different finite coupling strengths, thus having two different effective temperatures, can be in equilibrium with a common bath.  Thus a rudimentary message from this work is: for strong coupling thermodynamics one needs to separate the notion of  `equal temperature' (ET)  from `in equilibrium'(EQ).  EQ is a valid and, in our view,  a more fundamental notion than ET, which quantum thermodynamics at strong coupling, such as the zeroth law, should be based on. Only when the system-bath coupling  becomes vanishingly weak that temperature appearing in thermodynamic relations becomes universally defined and makes better physical sense. } 

We substantiate this claim with a description of our  methodology and rationale, as follows:
\begin{enumerate}[1)]
	\item For strong coupling thermodynamics a \textbf{nonequilibrium (NEq) effective temperature $T^*$} can be defined for Gaussian quantum systems \cite{NEqFE}. However, at equilibration with a heat bath of temperature $T_{\textsc{b}}$ , the system's effective temperature $T^*$ is not equal to, but (for the oscillator-field model studied) slightly higher than, that of the heat bath due to their non-negligible interaction. 
	\item Another identical system 2 with a \textbf{different coupling strength} albeit  interacting with an identical bath of the same bath temperature $T_{\textsc{b}}$ will have a different effective temperature $T^*_2$ at its equilibration.  Thus even when two identical systems 1 and  2 have reached equilibration with identical baths, if their coupling strengths are different,  their effective temperatures will be different. This implies that the zeroth law stated in terms of equal temperature does not hold for strong coupling thermodynamics.  
	%What remains is to investigate whether `in equilibrium' is a transitive relation. Namely, given two systems with different coupling strengths are separately in equilibrium with   the same bath, whether  they can be in equilibration with each other.  We first describe some earlier results which can be used for our present purpose. 
	\item \textbf{Equilibrium condition} can be defined for strongly coupled systems by examining the power transfer between the system and the bath.  Equilibrium condition is achieved when there is  no net energy flow between them such that power balance is reached. 
		\begin{enumerate}[(a)]
			\item Using the \textbf{power balance} criterion we have proven in earlier work that a chain of quantum harmonic oscillators when placed between two heat baths of different temperatures can indeed at late times approach a nonequilibrium steady state (NESS) \cite{AoP15}, and a fluctuation-dissipation relation for NESS can be established \cite{HH20b}. 
%Entanglement between two quantum oscillators in the same thermal bath has been studied in \cite{PLB15}. 
			\item For a system of $N$ quantum oscillators interacting strongly with and through a common quantum thermal field  {via bilinear coupling,} we have investigated the third law in terms of the heat capacity's temperature dependence near absolute zero {of the initial bath temperature}~\cite{QTD1}. We have also shown the existence of a  \textbf{generalized (tensorial) fluctuation-dissipation relation}  for oscillators at rest~\cite{QTD1} and for uniformly accelerating oscillators in the inertial vacuum~\cite{CPRPRD,CPRPLB}. 
			\item For a nonlinear oscillator bilinearly coupled to the thermal bath~\cite{FDRNESS}, the existence of a unique, stable equilibrium state implies a \textbf{nonperturbative} fluctuation-dissipation relation. 
	\end{enumerate}
	\item {\bf Present Work}. In Scenario (b) above, we assume the same coupling strength between each oscillator and the bath field. Here we shall focus on two quantum oscillators interacting strongly but with different strengths with the same thermal field bath.
What we need to do is to i) calculate their effective temperatures upon equilibration with identical baths, and show that they are different, and to ii) show that these two oscillators when placed in the same bath albeit at different effective temperatures can indeed reach equilibrium. 
%4) First law can be derived from the first and second law if the equilibrium condition can be reached by using energy (balance in the rate of energy flow) or entropy (reaching a maximum by some measure) respectively. 
	\item {\bf Equilibration $\neq$ Thermalization.} Here  in the context of thermodynamics one sees a clear distinction between \textit{in equilibrium} and \textit{in thermal equilibrium}. Thermalization demands more than equilibration. When a systems is coupled to a thermal bath, it may reach equilibrium, but not necessarily thermalize. Upon equilibration the reduced density matrix of the system does not have to assume the canonical distribution (thermal) form~\cite{GWT84,QTDpot} unless the coupling constant is vanishingly small. Only in this limit would the system be in thermal equilibrium with the bath and thermalization achieved. Without this added condition we can only say that the system is in equilibrium with the bath.
\end{enumerate}
%\end{description}

To establish notations we give in Sec.~II a short summary  of the newly introduced nonequilibrium effective temperature. Readers familiar with the results from~\cite{NEqFE} can skip over this section. In Sec.~III we delineate the implications from the dependence of the effective temperature on the system-bath coupling and why the zeroth law should not be phrased in terms of `equal temperature'. In Sec.~IV we show how how two oscillators equilibrate when they are coupled to a common bath with unequal but finite strengths. The existence of such dynamical equilibration allows us to show a set of generalized fluctuation-dissipation relations between these two systems, based on power balance.   In Sec.~V, we show why equilibrium condition is a more basic notion in the description of the zeroth law.

%%%%%%%%%%%%%%%%%%%%%%%%%%%%%%%%%%%%%%%%%%%%%%%%%

\section{Nonequilibrium effective temperature for a Gaussian system}\label{S:etbertw}

The concept of temperature is ill-defined for dynamically evolving nonequilibrium systems. In our recent work \cite{NEqFE} where an \textit{effective} temperature is defined, we did not force upon it. Rather, we attempted to  keep the definition of free energy  $\mathcal{F}_{\textsc{s}}(t)=\mathcal{U}_{\textsc{s}}(t)- T^{*} (t)\,\mathcal{S}_{vN}(t)$ intact for strong coupling thermodynamics, and examine each quantity in this relation in the nonequilibrium context. All of them have been studied and understood in varying degrees before,  except one, the temperature. The nonequilibrium free energy $\mathcal{F}_{\textsc{s}}$ of the system is related to the (closed-time-path) coarse-grained effective action, 
%\textcolor{red}{(strictly speaking, we write $\mathcal{F}_{\textsc{s}}$ as the effective action, but not derive $\mathcal{F}_{\textsc{s}}$ from the effective action because we have the difficulty to determine the appropriate normalization constant..)},
the von Neumann entropy $\mathcal{S}_{vN}$ is derived from the reduced density matrix of the system \cite{LH07,LH08} and the internal energy $\mathcal{U}_{\textsc{s}}$ from the Hamiltonian of mean force \cite{TH20,QTDpot,HMF}. The sought-after quantity we call NEq effective temperature $T^{*} (t)$ is  a time-dependent function which varies with the coupling strength. It assumes a familiar form only after the system equilibrates and acquires the proper physical meaning of the temperature of conventional (ultra-weak coupling) equilibrium thermodynamics.

This designer scheme is pretty attractive, but is the actual implementation feasible or tractable? That is the real challenge.
Arranging the expression of the density matrix operator into a (quasi-) canonical form~\cite{GWT84,HTL11a,KG10} leading to an explicit expression for the  effective temperature is not an easy task when non-commutativity of quantum operators is involved. Depending on how we rewrite the density matrix, there is no unique way to define an effective temperature.  
In addition, for a general system strongly coupled to the bath, the explicit form of its density matrix is usually not available, let alone a closed analytical form.

This seemingly impossible situation greatly improves for  the case of  Gaussian system linearly coupled to a Gaussian thermal bath. The implementation of this scheme and the identification of the effective temperature become feasible from these two observations: 1) If the initial states of the system and the bath are Gaussian, then the final (reduced) states of the system and the bath should also be Gaussian, and 2) the analytical expressions of the reduced density matrices of the Gaussian states are readily available. Thus, in the aforementioned example of a quantum Brownian oscillator strongly coupled to a Gaussian quantum thermal bath, even though the reduced density matrix operator of the oscillator is not of the Gibbsian form, its matrix elements are still Gaussian. 

The effective temperature of the oscillator is then identified when we compare the expressions of the density matrix elements in the coordinate representation with those for the thermal state of the oscillator. The effective temperature extracted this way should explicitly depend on the parameters of the oscillator system, in particular, the oscillator-bath coupling strength. Thus 
%An immediate consequence leads to the observation that 
when two oscillators are in contact with the same bath, if their coupling strengths are different, they have two different effective temperatures. This negates the `universal' nature,  and invalidates the transitivity relation  of temperature, as we  have gotten used to in conventional ultra-weak coupling thermodynamics. The adjective ``effective'' conveys these senses (beware of effective temperatures defined in other contexts under different assumptions). In what follows we will generalize the notion of effective temperature from an equilibrium setting to fully nonequilibrium conditions. We will do this for a time-evolving Gaussian quantum system strongly coupled to a Gaussian quantum thermal bath focusing on the consequences   of finite system-bath coupling for the zeroth law of thermodynamics.

For definiteness and mathematical simplicity, we consider a simple harmonic oscillator as an example of a Gaussian system $(S)$, initially in any Gaussian state,  coupled to a quantum scalar field, initially in a thermal state at   temperature $\beta_{\textsc{b}}^{-1}$, acting as the thermal bath $(B)$. Since such an initial configuration is in general not in equilibrium, their mutual interaction will induce subsequent evolution of both parties involved. The bath considered here has many more degrees of freedom and only a small fraction of the bath modes are expected to effectively interact with the oscillator system even when the oscillator-bath coupling is finite, not ultra-weak. The back-action from the system to the bath also tends to be negligible at late times~\cite{BSF00}, thus we can assume that the bath practically remains in the same initial thermal state.  This will no longer be the case for a finite-size bath or if the numbers of degrees of freedom of the system and of the bath are comparable. In the configuration we are considering, however, the corresponding back-reaction to the system can induce a substantial change and forces the system to evolve away from its initial configuration. The one important invariant is that throughout the evolution, the reduced density matrix of the system, although time dependent, remains Gaussian. It can be parametrized by an inverse temperature-like parameter $\vartheta$, the coherence parameter $\alpha$, the squeeze parameter $\zeta\in\mathbb{C}$ and the rotation angle $\theta\in\mathbb{R}$ in a form like~\cite{OL12,AA89,RW67,GA95,CD03,MC32a,MC32b}
\begin{equation}\label{E:bjdfhrse}
	\hat{\rho}_{\textsc{s}}=\hat{D}(\alpha)\hat{S}(\zeta)\hat{R}(\theta)\,\hat{\rho}_{\vartheta}\,\hat{R}^{\dagger}(\theta)\hat{S}^{\dagger}(\zeta)\hat{D}^{\dagger}(\alpha)\,.
\end{equation}
Here $\hat{D}(\alpha)$, $\hat{S}(\zeta)$ and $\hat{R}(\theta)$ are respectively the displacement, squeeze and the rotation operators
\begin{align*}
	\hat{D}(\alpha)&=\exp\biggl[\alpha\,\hat{a}^{\dagger}-\alpha^{*}\,\hat{a}\biggr]\,,&\hat{S}(\zeta)&=\exp\biggl[\frac{1}{2}\,\zeta^{*}\hat{a}^{2}-\frac{1}{2}\,\zeta\,\hat{a}^{\dagger\,2}\biggr]\,,&\hat{R}(\theta)&=\exp\biggl[-i\,\theta \Bigl(\hat{a}^{\dagger}\hat{a}+\frac{1}{2}\Bigr)\biggr]\,,
\end{align*}
with the annihilation and creation operators $a$, $a^{\dagger}$ satisfying the standard canonical commutation relation $[\hat{a}, \hat{a}^{\dagger}]=1$. The operator $\hat{\rho}_{\vartheta}$ takes on the canonical form
\begin{align}\label{E:bgkgere}
	\hat{\rho}_{\vartheta}&=\frac{1}{\mathcal{Z}_{\vartheta}}\,\exp\biggl[-\vartheta\,\Bigl(\hat{a}^{\dagger}\hat{a}+\frac{1}{2}\Bigr)\biggr]\,,&\mathcal{Z}_{\vartheta}&=\operatorname{Tr}\exp\biggl[-\vartheta\,\Bigl(\hat{a}^{\dagger}\hat{a}+\frac{1}{2}\Bigr)\biggr]=\frac{1}{2\sinh\frac{\vartheta}{2}}\,,
\end{align}
but beware that the positive real parameter $\vartheta$ should not outright be identified as an inverse temperature. The arbitrary parameters $\alpha$, $\zeta$, $\theta$, and $\vartheta$  will be determined by the system dynamics at the specific moments of time, expressible in terms the covariance matrix elements of the system. These parameters  are all functions of time.

We observe in \eqref{E:bjdfhrse} that by the cyclic property of the trace,
\begin{equation}
	\operatorname{Tr}\biggl\{\hat{D}(\alpha)\hat{S}(\zeta)\hat{R}(\theta)\,\exp\biggl[-\vartheta\,\Bigl(\hat{a}^{\dagger}\hat{a}+\frac{1}{2}\Bigr)\biggr]\,\hat{R}^{\dagger}(\theta)\hat{S}^{\dagger}(\zeta)\hat{D}^{\dagger}(\alpha)\biggr\}
\end{equation}
is actually given by $\mathcal{Z}_{\vartheta}$. It thus implies that $\mathcal{Z}_{\vartheta}$ will be our candidate for the nonequilibrium partition function $\mathcal{Z}_{\textsc{s}}$ for the reduced density matrix $\hat{\rho}_{\textsc{s}}$ of the oscillator system at any arbitrary moment. After some algebraic manipulations~\cite{NEqFE,RW67,GA95,MC32a,MC32b}, we find this nonequilibrium partition function $\mathcal{Z}_{\textsc{s}}$ can be written as
\begin{align}\label{E:dgjherhs}
	\mathcal{Z}_{\textsc{s}}(t)=\frac{1}{2}\,\operatorname{csch}\frac{\vartheta(t)}{2}=\Bigl(ab-c^{2}-\frac{1}{4}\Bigr)^{\frac{1}{2}}\,,
\end{align}
with $b(t)=\langle\hat{\chi}^{2}(t)\rangle$, $a(t)=\langle\hat{p}^{2}(t)\rangle$, and $c(t)=\dfrac{1}{2}\langle\bigl\{\hat{\chi}(t),\,\hat{p}(t)\bigr\}\rangle$, the covariance matrix elements of the oscillator. These expectation values are taken with respect to the density matrix $\hat{\rho}_{\textsc{s}}(t)$. The expression inside the large parentheses is connected to the Robertson-Schr\"odinger uncertainty relation
\begin{equation}
	\langle\hat{\chi}^{2}\rangle\langle\hat{p}^{2}\rangle-\dfrac{1}{4}\langle\bigl\{\hat{\chi},\,\hat{p}\bigr\}\rangle^{2}\geq\frac{1}{4}\,.
\end{equation}
It is interesting to note that the uncertainty principle was originally derived based on the general consideration of non-commutativity of operators, so it does not specifically rely on the (non)equilibrium property of the reduced (non)Gaussian quantum system. Its appearance may be linked to  the fact that the dynamics of a Gaussian system can be completely and uniquely determined by its covariance matrix elements, and the uncertainty principle is embedded in them. The expression on the righthand side of \eqref{E:dgjherhs} will play a significant role in the nonequilibrium thermodynamics of the Gaussian systems.

\begin{figure}
\centering
    \scalebox{0.35}{\includegraphics{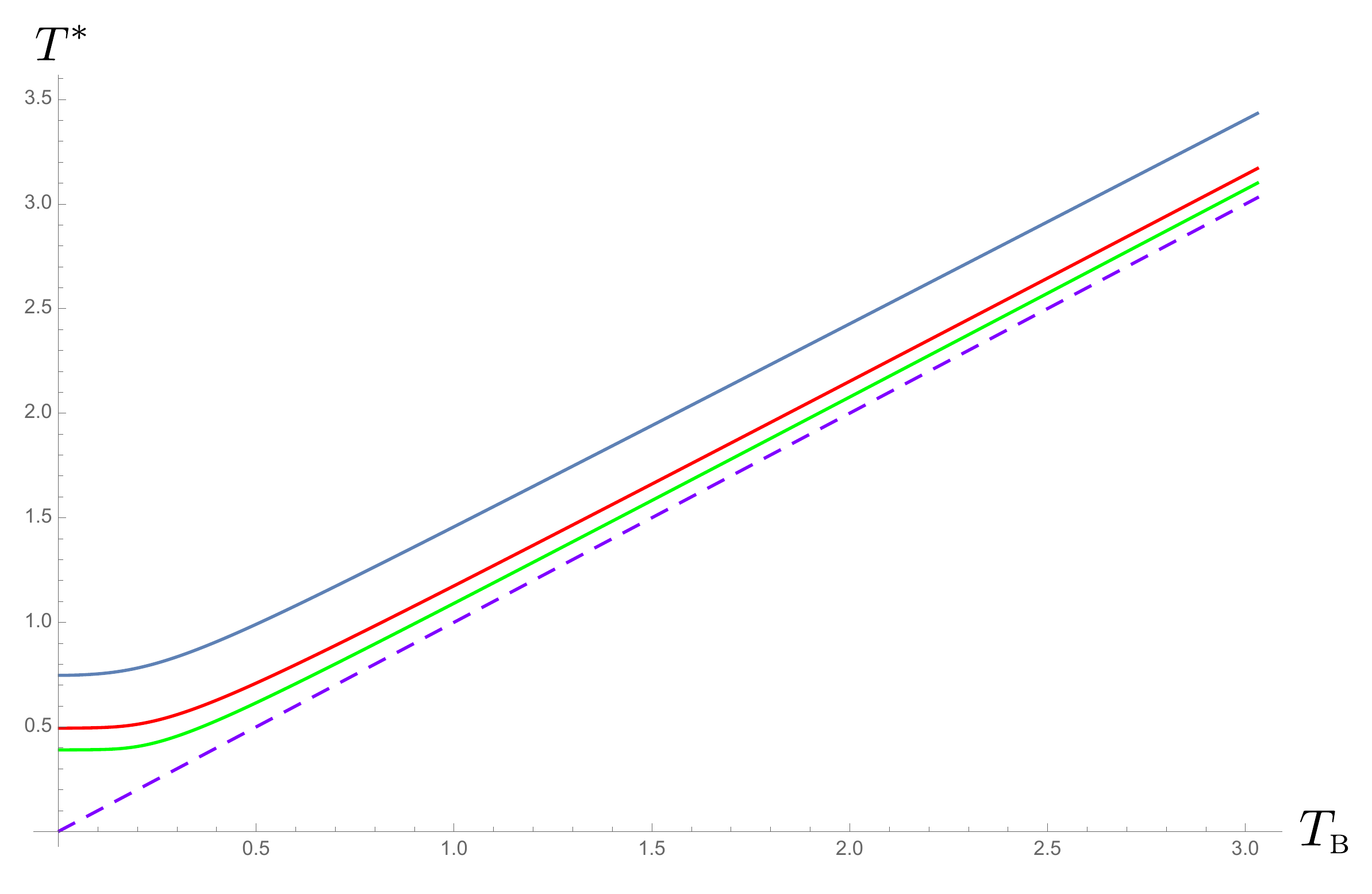}}
    \caption{The dependence of the effective temperature $T^{*}=\beta_{\textsc{eff}}^{-1}$ of the system on the initial temperature $T_{\textsc{b}}=\beta_{\textsc{b}}^{-1}$ of the bath. The parameters are normalized with respect to the resonance frequency $\Omega=\sqrt{\omega_{\textsc{r}}^{2}-\gamma^{2}}$ such that $m=1\,\Omega$, and the cutoff frequency $\Lambda=1000\,\Omega$. We choose the damping constant $\gamma=0.3$ for the blue solid curve, $\gamma=0.1$ for the red solid curve, and $\gamma=0.03$ for the green solid curve. The evolution time $t$ is long enough with $\gamma\,t=3.6$, such that the dynamics is sufficiently relaxed. Thus the curves become independent of the initial conditions of the system. The dashed line is a reference line representing the bath's initial temperature.}\label{Fi:effectiveT}
\end{figure}
Eq.~\eqref{E:dgjherhs} has identical appearance to the equilibrium partition function we are familiar with in the conventional, weak-coupling thermodynamics, except for the time dependence. This opens the way for us to identify the effective temperature $T^{*}(t)=\beta_{\textsc{eff}}^{-1}(t)$ of the system in the duration of the nonequilibrium evolution by
\begin{equation}
	\vartheta(t)=\beta_{\textsc{eff}}(t)\,\omega_{\textsc{r}}\,,
\end{equation}
where $\omega_{\textsc{r}}$ is the physical frequency of the oscillator. In terms of the covariance matrix elements $a$, $b$ and $c$, this effective inverse temperature is given by
\begin{equation}\label{E:gbksjhd}
	\beta_{\textsc{eff}}(t)=\frac{2}{\omega_{\textsc{r}}}\,\ln\frac{1+\sqrt{1+4\mathfrak{S}(t)}}{2\sqrt{\mathfrak{S}(t)}}\,,
\end{equation}
where
\begin{align}
	\mathfrak{S}(t)=a(t)b(t)-c^{2}(t)-\frac{1}{4}\geq0\,,
\end{align}
gives the the Robertson-Schr\"odinger uncertainty relation at any moment. Although the covariance matrix elements are defined with respect to the reduced density matrix operator $\hat{\rho}_{\textsc{s}}(t)$, they can alternatively be found with the help of the quantum Langevin equation~\cite{NEqFE} that describes the reduced dynamics of the oscillator system under the influence of the bath.

It is known that, independent of its initial state, the reduced system we consider here will eventually relax to an equilibrium state at time scale much greater than $\gamma^{-1}$, the damping constant, which contains the coupling strength between the oscillator system and the thermal bath and quantifies its dissipative back-reaction on the system dynamics. In this final equilibrium state all the physical observables of the system, in particular, the covariance matrix elements, will become time-independent, but the corresponding density matrix operator does not take on the thermal form unless $\gamma$ approaches zero. We expect the effective temperature defined in \eqref{E:gbksjhd} to have similar dynamical features as any other system observables, and will relax to a constant upon equilibration. An effective temperature  defined in this way thus reflects more the dynamical characteristics of the system than its statistical nature. When the system relaxes to an equilibrium state it will begin to acquire the statistical properties  passed on from the thermal bath, but is skewed by the finite system-bath coupling strength. This also explains why in the limit of vanishing coupling, the system adheres to the thermal statistics at the bath's temperature.

\section{zeroth law not in terms of equal temperature}

\begin{figure}
\centering
    \scalebox{0.55}{\includegraphics{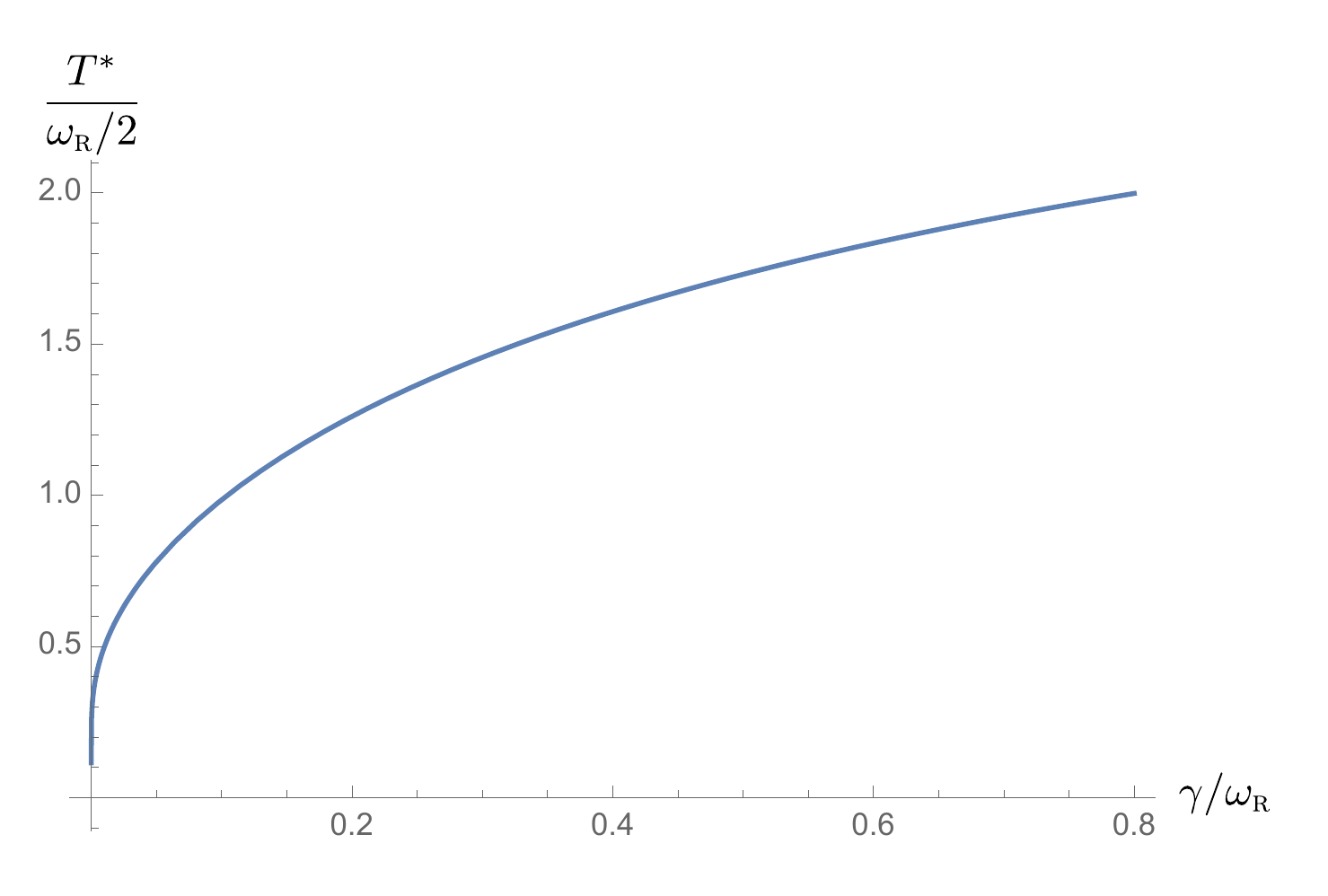}}
    \caption{The dependence of the effective temperature $T^{*}=\beta_{\textsc{eff}}^{-1}$ of the reduced system in the final equilibrium state on the system-bath coupling constant $\gamma$. We focus on the case where the initial temperature of the bath is zero. Here we choose $m=1\,\omega_{\textsc{r}}$, and the cutoff frequency $\Lambda=1000\,\omega_{\textsc{r}}$. Note how the curve rapidly returns to $T^{*}\to0$ in the neighborhood of $\gamma\simeq0$.}\label{Fi:effectivT-1}
\end{figure}

The effective temperature introduced in the previous section in general is real and non-negative in the final equilibrium state due to the Robertson-Schr\"odinger uncertainty principle. In fact it is always greater than zero. Specially engineered states of the system may take exceptions, such as a  zero-temperature thermal state. However this is highly unlikely if the system can be let go through unrestrained nonequilibrium relaxation allowing only interaction with the bath. Even if the bath is initially set at zero temperature, the resulting effective temperature in the final equilibrium state is still positive and the deviation from the bath temperature grows with the coupling strength, as shown in Fig.~\ref{Fi:effectivT-1}. Generically, the effective temperature will be higher than the initial bath temperature, if the Gaussian system couples more strongly to the thermal bath. Only in the limiting cases that 1) the system-bath coupling is vanishingly small, or 2) the initial temperature of the bath is sufficiently high, will the effective temperature of the system approach the bath temperature,  acquiring the status of the absolute temperature. The greatest deviation of the effective temperature from the initial bath temperature will appear at the low bath temperature  and strong coupling regimes, as shown in Fig.~\ref{Fi:effectiveT}. This is a particularly important property of the effective temperature in the context of the zeroth law. 

Consider that we prepare two identical harmonic oscillators $1$, $2$ with the same physical parameters like mass and oscillating frequency except for their coupling strengths, which we denote as $\gamma_{\textsc{1}} \neq \gamma_{\textsc{2}}$.  We also prepare two separate but macroscopically identical  thermal baths $B$, with the same temperature $T_{\textsc{b}}=\beta_{\textsc{b}}^{-1}$. 
%This example may be a little extreme because subsystem $B$ has a much greater number of degrees of freedom than the other two subsystems. However, this may be the simplest configuration to illustrate the issues of the zeroth law at finite coupling. 
Next we let oscillator $1$ couple to its bath $B$ and wait for a sufficiently long time until they reach equilibration. From the final reduced density matrix operator, we can identify the effective temperature of oscillator $1$ to be $T_{1}^{*}$.  We carry out the same protocol for oscillator $2$.  Wait for oscillator $2$ to equilibrate with its bath $B$. Because of the different coupling strength, it will have a different relaxation time scale, and after equilibration we will find that oscillator $2$ has a different effective temperature $T_{2}^{*}$. From Figs.~\ref{Fi:effectiveT}, ~\ref{Fi:effectivT-1} and the previous discussions, we establish that
\begin{equation}
	T_{1}^{*}\neq T_{2}^{*}\,.
\end{equation}
Furthermore neither of them is equal to $T_{\textsc{b}}$, even though in the end oscillator $1$ is in equilibrium with its bath $B$, and oscillator $2$ is in equilibrium with its identical bath $B$. The final state of oscillator $1$ will be different from that of oscillator $2$ while the states of the two identical baths $B$'s are practically the same, as their initial states are equal by our preparation, and by virtue of their status as a bath, having an overwhelmingly larger number of degrees of freedom than either oscillator~\cite{BSF00}.

Two comments about finding the effective temperature of a system: 1) As discussed earlier, the values of the effective temperatures of oscillators $1$, $2$ at equilibration will be independent of their initial states, so in principle we do not need to prepare oscillators $1$ and $2$ in the same initial state. 2) Since the effective temperature is expressed in terms of expectation values, i.e., the covariance matrix elements $a$, $b$ and $c$, it is understood as an averaged value. It does not correspond to a particular value for a specific realization of the system-bath configuration permissible by the governing quantum probability distribution.  Like finding the expectation value of any system observable, we only need to prepare a sufficiently large ensemble of such open systems. This protocol applies to the effective temperature of the system.

With this explanation we begin to see the looming trouble of casting the zero law at strong coupling in terms of equal temperature. When subsystem $1$ is in contact with its bath $B$, after they reach equilibration, its effective temperature $T_{1}^{*}$ is different from the effective temperature $T_{2}^{*}$ of
%@ maybe better to leave this out for now since it alludes to ETH etc \footnote{here we put aside the issues how a thermometer reads off the temperature of a system $A$. it is interesting to understand how a thermometer measures the temperature of a equilibrated but not thermalized system. An involved subtle issue is that if the thermometer is a macroscopic system, then there may be additional mechanism such as ETH to thermalize. thus it is beyond the scape of this paper.}. 
subsystems $2$ after it has reached equilibration with its identical bath $B$. Moreover, even within each subsystem and its bath, their  temperatures at equilibrium $T_{1}^{*}$, $T_{\textsc{b}}$ are different. Thus equal temperature is not a good criterion for the zeroth law of thermodynamics at strong coupling. %Each oscillator has reached equilibrium with its $B$ yet their effective temperatures are different. 
 
% @Let's group this in the next section about equilibrium and just focus on temperature here @ This has invalidated the starting statement in the zeroth law that \textit{if two thermodynamic systems are each in thermal equilibrium with a third one}. Strictly speaking, at finite system-bath coupling, after subsystem $1$ relaxes to equilibrium, although it does not have net energy exchange with subsystem $B$, its reduced density matrix does not take the canonical form. Thus we will say subsystems $1$ and $B$ are in equilibrium, instead of the stronger condition of thermal equilibrium. Secondly, they do not have the same temperature, so a common rephrasing of the zeroth law in terms of temperature, as a tag of the system in thermal equilibrium, becomes futile. Thus it seems sensible to tone down the aforementioned statement to a weaker form that \textit{if two thermodynamic systems are each in equilibrium with a third one} without literally referencing to thermal equilibrium and the accompanying notion of temperature @.

\section{zeroth law in terms of equilibrium conditions}

We now explore the ``in equilibrium" conditions more closely. Let subsystems $1$ and $2$ be simultaneously coupled to a common subsystem $B$ with different but finite coupling strengths, assuming no direct interaction between subsystems $1$ and $2$. The special cases of \textit{equal coupling strength} have been investigated in~\cite{SY09, PRD15,QTD1,CPRPRD,CPRPLB} in the context of thermal entanglement and equilibration of nonequilibrium evolutions. It has been shown that if subsystem $B$ is initially in its thermal state, but subsystems $1$ and $2$ are in any Gaussian state, then even though the entire system starts out with a nonequilibrium configuration, each subsystem will equilibrate at later times when there is no net energy flow between each pair of subsystems, as dictated by the fluctuation-dissipation and the correlation-propagation relations~\cite{QTD1,CPRPRD,CPRPLB}. The only prerequisites are that subsystem $1$ and $2$ are not too close in proximity and their coupling strengths with subsystem $B$ are not too large~\cite{PRD15} because these cases tend to induced dynamical instability to the assumed configuration. Thus subsystems $1$ and $2$ and $B$ will come in equilibrium at late times. Since the degrees of freedom of subsystem $B$ are infinitely larger the degrees of freedom of subsystem $1$ and $2$ then, following the arguments in~\cite{BSF00}, subsystem $B$ practically remains at its initial temperature. But, due to the finite coupling between $1-B$ and $2-B$, it is reasonable to project that the subsystems $1$ and $2$ will not reach an effective temperature that has the same value as the initial temperature of subsystem $B$.

\subsection{Two uncoupled oscillators with unequal oscillator-bath couplings}\label{S:gbrh}
 
We now consider the more general  configuration of \textit{unequal oscillator-bath couplings}. We will show that they can still equilibrate at sufficiently late times, and after equilibration, these three subsystems in general can have three distinct effective temperatures.

To be specific, we consider two mutually uncoupled harmonic oscillators 1, 2, whose displacements are denoted by operators $\hat{\chi}_{1}$, $\hat{\chi}_{2}$. They are simultaneously coupled to a common quantum field $\hat{\phi}$ with different coupling strengths $g_{1}$, $g_{2}$ respectively. The field is initially prepared in a thermal state at temperature $\beta_{\textsc{b}}^{-1}$, and acts as the thermal bath. The equations of motion for the oscillator system take the form
\begin{align}\label{E:eirieiiree}
	m\,\ddot{\hat{\bm{\Xi}}}(t)+m\,\bm{\Omega}^{2}_{\textsc{b}}\,\hat{\bm{\Xi}}(t)-\int_{0}^{t}\!ds\;\mathbf{g}^{T}\cdot\mathbf{G}_{R,0}^{(\phi)}(t-s)\cdot\mathbf{g}\cdot\hat{\bm{\Xi}}(s)=\mathbf{g}^{T}\cdot\hat{\bm{\Phi}}_{h}(t)\,,
\end{align}
with $t>0$. Here $\hat{\bm{\Xi}}=(\hat{\chi}_{1},\hat{\chi}_{2})^{T}$ is the column vector, representing the displacement operators of two oscillators. The superscript $T$ denotes a matrix transpose. The diagonal matrix 
\begin{align}
	\bm{\Omega}^{2}_{\textsc{b}}=\begin{bmatrix}\omega_{b_{1}}^{2}&0\\0&\omega_{b_{2}}^{2}\end{bmatrix}
\end{align}
gives the bare frequencies of two oscillators. They take on different values, but after frequency renormalization, they will be assumed to have the same physical frequency $\omega_{\textsc{r}}$. The diagonal elements of the  matrix $\mathbf{g}=\operatorname{diag}(g_{1},g_{2})$ give the respective coupling strengths of the oscillators with the bath. The noise force from the bath is described by the column vector $\hat{\bm{\Phi}}_{h}$,
\begin{equation}
	\hat{\bm{\Phi}}_{h}(t)=\begin{bmatrix}\hat{\phi}_{h}(\mathbf{z}_{1},t)\\\hat{\phi}_{h}(\mathbf{z}_{2},t)\end{bmatrix}\,,
\end{equation}
where $\hat{\phi}_{h}(\mathbf{x},t)$ is the free massless scalar Klein-Gordon field operator, and $\mathbf{z}_{i}$ is the spatial location of oscillator $i=1$, 2. Thus the amplitudes of the driving forces on the two oscillators are different but correlated. The retarded Green function matrix $\mathbf{G}_{R,0}^{(\phi)}(t-s)$ of the free scalar field inside the nonlocal expression of the equation of motion is defined by
\begin{equation}
	\mathbf{G}_{R,0}^{(\phi)}(t-s)=i\,\theta(t-s)\,\operatorname{Tr}_{\phi}\Bigl(\hat{\rho}_{\beta_{\textsc{b}}}^{(\phi)}\bigl[\hat{\bm{\Phi}}_{h}^{\vphantom{T}}(t),\,\hat{\bm{\Phi}}_{h}^{T}(s)\bigr]\Bigr)\,,
\end{equation}
such that $[\mathbf{G}_{R,0}^{(\phi)}(t-s)]_{ij}=G_{R,0}^{(\phi)}(\mathbf{z}_{i},t;\mathbf{z}_{j},s)$. The density matrix $\hat{\rho}_{\beta_{\textsc{b}}}^{(\phi)}$ is the initial thermal state of the free field at temperature $\beta_{\textsc{b}}^{-1}$.

The solution to the equations of motion can be obtained by applying the Laplace transformation to Eq.~\eqref{E:eirieiiree}
\begin{align}
	\Bigl\{m\bigl(z^{2}\mathbf{I}+\bm{\Omega}^{2}_{\textsc{b}}\bigr)-\mathbf{g}^{T}\cdot\widetilde{\mathbf{G}}_{R,0}^{(\phi)}(z)\cdot\mathbf{g}\Bigr\}\cdot\hat{\widetilde{\bm{\Xi}}}(z)=mz\,\hat{\bm{\Xi}}(0)+m\,\dot{\hat{\bm{\Xi}}}(0)+\mathbf{g}^{T}\cdot\hat{\widetilde{\bm{\Phi}}}_{h}(z)\,,
\end{align}
where the Laplace transformation of a function $f(t)$ is defined by
\begin{equation}
	\widetilde{f}(z)=\int_{0}^{\infty}\!dt\;f(t)\,e^{-zt}\,,
\end{equation}
with $\operatorname{Re}z>0$. Solving for $\hat{\widetilde{\bm{\Xi}}}(z)$ gives
\begin{equation}
	\hat{\widetilde{\bm{\Xi}}}(z)=\widetilde{\mathbf{G}}_{R}^{(\chi)}(z)\cdot\Bigl[mz\,\hat{\bm{\Xi}}(0)+m\,\dot{\hat{\bm{\Xi}}}(0)\Bigr]+\widetilde{\mathbf{G}}_{R}^{(\chi)}(z)\cdot\mathbf{g}^{T}\cdot\hat{\widetilde{\bm{\Phi}}}_{h}(z)\,,
\end{equation}
where $\widetilde{\mathbf{G}}_{R}^{(\chi)}(z)$ is the Laplace transform of the retarded Green's function of the interacting oscillators
\begin{equation}\label{E:ngbshkdjf}
	\widetilde{\mathbf{G}}_{R}^{(\chi)}(z)=\Bigl\{m\bigl(z^{2}\mathbf{I}+\bm{\Omega}^{2}_{\textsc{b}}\bigr)-\mathbf{g}^{T}\cdot\widetilde{\mathbf{G}}_{R,0}^{(\phi)}(z)\cdot\mathbf{g}\Bigr\}^{-1}\,.
\end{equation}
The solution $\hat{\bm{\Xi}}(t)$ to Eq.~\eqref{E:eirieiiree} is then given by performing the inverse Laplace transformation
\begin{equation}
	\hat{\bm{\Xi}}(t)=\int_{C}\!\frac{dz}{2\pi i}\;\widetilde{\bm{\Xi}}(z)\,e^{zt}\,,
\end{equation}
where the closed contour $C$ encloses all the poles of $\widetilde{\mathbf{G}}_{R}^{(\chi)}(z)$ on the complex $z$ plane.

Suppose the poles occur at $z_{p}$, then we want $\operatorname{Re}z_{p}<0$ to ensure that $\mathbf{G}_{R}^{(\chi)}(t)$ exponentially decays with time and the late-time dynamics of the system will be independent of the initial conditions. To see what conditions will fulfill this requirement, we explicitly examine one of the components of \eqref{E:eirieiiree}. We have, say for oscillator 1, the secular equation whose zeroes will tell the locations of the poles, given by
\begin{equation}\label{E:bfdjewr} 
	z^{2}+2\gamma_{1}z+\omega^{2}_{\textsc{r}}-\frac{g_{1}g_{2}}{4\pi m}\,\frac{e^{-zd}}{d}=0\,,
\end{equation}
where $d=\lvert\mathbf{z}_{1}-\mathbf{z}_{2}\rvert$ is the fixed separation between the two oscillators, and the damping constant $\gamma_{1}=g_{1}^{2}/8\pi m$ for oscillator 1 will determine its relaxation time scale. The fourth term in \eqref{E:bfdjewr} depicts the non-Markovian effects between the two oscillators mediated by the common bath field. The symmetric arrangement of $g_{1}$, $g_{2}$ shows that both oscillators will experience the same mutual effect, and the presence of $e^{-zd}$ indicates that this non-Markovian effect is not instantaneous; it will take finite time to propagate from one oscillator to the other. A similar but more complicated structure also appears when the bath field is described by a quantized electromagnetic field~\cite{HL06}. The renormalized frequency $\omega_{\textsc{r}}$ is related to the bare frequency $\omega_{\textsc{b}_{1}}$ of oscillator 1 by
\begin{align}\label{E:gkshesdfs}
	\omega_{\textsc{r}}^{2}=\omega_{\textsc{b}_{1}}^{2}-\frac{g_{1}^{2}}{2\pi^{2}m\varepsilon}\,,
\end{align}
where $\varepsilon$ is the shortest length scale in the model, and is usually determined by the cutoff frequency $\Lambda$ by $\varepsilon\sim\Lambda^{-1}$. As mentioned earlier, we choose the bare frequency $\omega_{\textsc{b}_{1,2}}$ and $\gamma_{1,2}$ such that both oscillators have the same renormalized, physical frequency $\omega_{\textsc{r}}$.

Let us examine the large separation and short range behaviors. For large $d$, the term that accounts for the non-Markovian effect is subdominant, so we can safely say $\operatorname{Re}z_{p}<0$ for this case. For small $d$, we may approximate  \eqref{E:bfdjewr} as
\begin{equation}\label{E:bfdjewr2} 
	z^{2}+2\gamma_{1}z+\omega^{2}_{\textsc{r}}-\frac{g_{1}g_{2}}{4\pi md}\,\bigl(1-zd\bigr)=z^{2}+\Bigl(\frac{g_{1}^{2}}{4\pi m}+\frac{g_{1}g_{2}}{4\pi m}\Bigr)\,z+\Bigl(\omega^{2}_{\textsc{r}}-\frac{g_{1}g_{2}}{4\pi md}\Bigr)=0\,.
\end{equation}
This shows that when the oscillators' separation $d$ is too small, or $g_{1}g_{2}$ is too large, the effective oscillating frequency squared, the last pair of parentheses in \eqref{E:bfdjewr2},  can be negative. It may turn a harmonic oscillator into an inverted oscillator with possible dynamic instability. From  the solution to \eqref{E:bfdjewr2},
\begin{align}
	z_{p}=-\Bigl(\frac{g_{1}^{2}}{4\pi m}+\frac{g_{1}g_{2}}{8\pi m}\Bigr)\pm i\biggl[\Bigl(\omega^{2}_{\textsc{r}}-\frac{g_{1}g_{2}}{4\pi md}\Bigr)-\Bigl(\frac{g_{1}^{2}}{4\pi m}+\frac{g_{1}g_{2}}{8\pi m}\Bigr)\biggr]^{\frac{1}{2}}\,,
\end{align}
we immediately see that if
\begin{equation}
	\omega^{2}_{\textsc{r}}-\frac{g_{1}g_{2}}{4\pi md}>0\,,
\end{equation}
then $\operatorname{Re}z_{p}$ is always negative, and the oscillator will have an attractive fixed point. Otherwise, we may find that $\operatorname{Re}z_{p}$ can be positive, so we have unstable, runaway dynamics. From this, to ensure well behaved oscillator dynamics, we require that the separation between the two oscillators should not be too small, and that the product of the coupling constants $g_{1}g_{2}$ should not be too large, so that the square of the effective oscillating frequency to have a positive value. This will allow the oscillators-field system to reach equilibrium at late times.

\subsection{Energy balance and steady state at equilibration}\label{S:reksd}

Next we turn to the energy balance in the system-bath upon the system's equilibration. The existence of a final equilibrium steady state guarantees that there is neither net energy exchanges between oscillators, nor between each oscillator and the bath field. This condition makes way for  a generalized fluctuation-dissipation relation (FDR) to exist. The existence proofs for $N$ quantum oscillators interacting with a common quantum field with equal coupling strengths have been given in our earlier work \cite{QTD1,CPRPRD,CPRPLB}. Here, to answer the questions posed earlier concerning the zeroth law at strong coupling we need to examine whether the generalized fluctuation-dissipation relation still holds if the two oscillators are coupled to the thermal bath with \textit{different} strengths. Here we only use two oscillators for illustration. But our arguments can be generalized to any number of oscillators. This is an extension of previous results in~\cite{QTD1}.

We first decompose the equation of motion for oscillator 1 into
\begin{align}\label{E:gnkjsgsg}
	m\,\ddot{\hat{\chi}}_{1}(t)+m\omega_{\textsc{r}}^{2}\,\hat{\chi}_{1}(t)&+g_{1}^{2}\Gamma^{(\phi)}(t)\,\hat{\chi}_{1}(0)\\
	&=g_{1}\hat{\phi}_{h}(\mathbf{z}_{1},t)-g_{1}^{2}\int_{0}^{t}\!ds\;\Gamma^{(\phi)}(t-s)\dot{\hat{\chi}}_{1}(s)+g_{1}g_{2}\int_{0}^{t}\!ds\;G_{R,0}^{(\phi)}(\mathbf{z}_{1},t;\mathbf{z}_{2},s)\hat{\chi}_{2}(s)\,.\notag
\end{align}
where we have introduced $G_{R,0}^{(\phi)}(\mathbf{z}_{1},t;\mathbf{z}_{1},s)=-d\Gamma^{(\phi)}(t-s)/dt$,
\begin{align}\label{E:gbsjhssd}
	\omega_{\textsc{r}}^{2}&=\omega_{\textsc{b}_{1}}^{2}-\frac{g_{1}^{2}}{m}\,\Gamma^{(\phi)}(0)\,.
\end{align}
For the scalar field bath, the function $\Gamma^{(\phi)}(t)$ is proportional to a delta function $\delta(t)$. When we introduce a cutoff length scale $\varepsilon$, Eq.~\eqref{E:gbsjhssd} will be written into \eqref{E:gkshesdfs}. Eq.~\eqref{E:gnkjsgsg} contains a sudden kick at the initial position of the oscillator. This is a consequence that there is no appropriate length scale in the initial configuration we assume~\cite{FRH11}. In more realistic settings, there always exists some initial correlation between the system and the bath, or that the system-bath coupling takes a finite time to take effect. Thus  the sudden kick is not so drastic. It will be replaced by a smoother transition over some time scale of the aforementioned processes. Either way, its effect will not be relevant to our system's late-time dynamics, so we will ignore it in the following discussion.

From the righthand side of \eqref{E:gnkjsgsg}, we observe that there are three channels through which a given oscillator will exchange energy with its surrounding, including its neighboring oscillator. 1) The local quantum field fluctuations will drive the oscillator, pumping energy from the field, in the same way as a driven oscillator by an external force. The main difference is that the noise associated with the field fluctuations will introduce random motion to the oscillator, and pass on their statistical characteristics to it. 2) The quantum dissipative   motion of the oscillator, as a consequence of the driving noise, will dissipate energy back to the surrounding field~\cite{AFM0}. These two channels explicitly depend only on the coupling constant $g_{1}$ of oscillator 1, and are local in nature. Finally, 3) The energy transfer between the two oscillators through the field which propagates their mutual influences. The appearance of $g_{1}$, $g_{2}$ highlights its nonlocal character. In addition, this part depends on the system's  evolutionary history, and is thus non-Markovian, and yet most important of all, causal.

Take the example of oscillator 1. The power $P_{\xi}$ pumped in by the bath field, the power lost by dissipation $P_{\gamma}$, and the energy exchange rate $P_{c}$ via the field-mediated non-Markovian effect are respectively expressed by~\cite{QTD1,HH20b}
\begin{align}
	P_{\xi}^{(1)}(t)&=g_{1}\int_{0}^{t}\!ds\;\frac{\partial}{\partial t}\Bigl[\mathbf{G}_{R}^{(\chi)}(t-s)\Bigr]{}_{1i}\,\mathbf{g}_{ij}\,\Bigl[\mathbf{G}_{H,0}^{(\phi)}(s,t)\Bigr]{}_{j1}\,,\label{E:dkhejr1}\\
	P_{\gamma}^{(1)}(t)&=-g_{1}^{2}\int_{0}^{t}\!ds\;\Gamma^{(\phi)}(t-s)\frac{\partial^{2}}{\partial t\,\partial s}\Bigl[\mathbf{G}_{H}^{(\chi)}(s,t)\Bigr]{}_{11}\,,\label{E:dkhejr2}\\
	P_{c}^{(1)}(t)&=g_{1}g_{2}\int_{0}^{t}\!ds\;\Bigl[\mathbf{G}_{R,0}^{(\phi)}(t-s)\Bigr]_{12}\frac{\partial}{\partial t}\Bigl[\mathbf{G}_{H}^{(\chi)}(s,t)\Bigr]{}_{21}\,,\label{E:dkhejr3}
\end{align}
where at times greater than the relaxation time $\gamma_{1}^{-1}$, the contribution from the initial conditions are exponentially suppressed. Here $\mathbf{G}_{H,0}^{(\phi)}(t,t')$, $\mathbf{G}_{H}^{(\chi)}(t,t')$ are the Hadamard functions respectively of the free field and the oscillator. For example, the former is defined by
\begin{equation}
	\mathbf{G}_{H,0}^{(\phi)}(t,t')=\frac{1}{2}\operatorname{Tr}_{\phi}\Bigl(\hat{\rho}_{\beta_{\textsc{b}}}^{(\phi)}\bigl\{\hat{\bm{\Phi}}_{h}^{\vphantom{T}}(t),\,\hat{\bm{\Phi}}_{h}^{T}(s)\bigr\}\Bigr)\,.
\end{equation}
Since for the linear, Gaussian system under study it has been shown~\cite{QTD1} that at late times the contributions from the nonstationary part of the Hadamard function will be exponentially smaller than those from the stationary part, we can write the Hadamard function, say, $\mathbf{G}_{H}^{(\chi)}(s,t)$ in \eqref{E:dkhejr2} and \eqref{E:dkhejr3} in a stationary form $\mathbf{G}_{H}^{(\chi)}(s-t)=\mathbf{G}_{H}^{(\chi)}(t-s)$. Considering the retarded nature of $\mathbf{G}_{R}^{(\chi)}(t-s)$, $\Gamma^{(\phi)}(t-s)$, and $\mathbf{G}_{R}^{(\chi)}(t-s)$, we thus find in the limit $t\to\infty$,  \eqref{E:dkhejr1}--\eqref{E:dkhejr3}  can be written as~\cite{QTD1,CPRPLB}
\begin{align}
	P_{\xi}^{(1)}(\infty)&=g_{1}\int_{-\infty}^{\infty}\!\frac{d\kappa}{2\pi}\;\kappa\,\Bigl[\operatorname{Im}\overline{\mathbf{G}}_{R}^{(\chi)}(\kappa)\Bigr]_{1i}\,\mathbf{g}_{ij}\,\Bigl[\overline{\mathbf{G}}_{H,0}^{(\phi)}(\kappa)\Bigr]_{j1}\,,\\
	P_{\gamma}^{(1)}(\infty)&=-g_{1}^{2}\int_{-\infty}^{\infty}\!\frac{d\kappa}{2\pi}\;\kappa\,\Bigl[\operatorname{Im}\overline{\mathbf{G}}_{R,0}^{(\phi)}(\kappa)\Bigr]{}_{11}\Bigl[\overline{\mathbf{G}}_{H}^{(\chi)}(\kappa)\Bigr]{}_{11}\,,\\
	P_{\gamma}^{(1)}(\infty)&=-g_{1}g_{2}\int_{-\infty}^{\infty}\!\frac{d\kappa}{2\pi}\;\kappa\,\Bigl[\operatorname{Im}\overline{\mathbf{G}}_{R,0}^{(\phi)}(\kappa)\Bigr]{}_{12}\Bigl[\overline{\mathbf{G}}_{H}^{(\chi)}(\kappa)\Bigr]{}_{21}\,,
\end{align}
where we have used the facts that 1) $\operatorname{Im}\overline{\mathbf{G}}_{R}(\kappa)$ is an odd function of $\kappa$, 2) $\overline{\mathbf{G}}_{H}(\kappa)$ is an even function of $\kappa$, and 3) $\overline{\mathbf{G}}_{R,0}^{(\phi)}(\kappa)=i\kappa\,\overline{\Gamma}^{(\phi)}(\kappa)$. The Fourier transformation $\overline{f}(\kappa)$ of a function $f(t)$ is defined by
\begin{equation}
	\overline{f}(\kappa)=\int_{-\infty}^{\infty}\!dt\;f(t)\,e^{i\kappa t}\,.
\end{equation}

In the steady state at late times, there is no net energy exchange between the subsystems, that is,
\begin{equation}
	P_{\xi}^{(1)}(\infty)+P_{\gamma}^{(1)}(\infty)+P_{\gamma}^{(1)}(\infty)=0\,,
\end{equation}
which in turn  requires
\begin{equation}\label{E:rjhvjgssg}
	\int_{-\infty}^{\infty}\!\frac{d\kappa}{2\pi}\;\kappa\biggl\{\Bigl[\operatorname{Im}\overline{\mathbf{G}}_{R}^{(\chi)}(\kappa)\Bigr]_{1i}\,\mathbf{g}_{ij}\,\Bigl[\overline{\mathbf{G}}_{H,0}^{(\phi)}(\kappa)\Bigr]_{j1}-\Bigl[\operatorname{Im}\overline{\mathbf{G}}_{R,0}^{(\phi)}(\kappa)\Bigr]{}_{1i}\,\mathbf{g}_{ij}\,\Bigl[\overline{\mathbf{G}}_{H}^{(\chi)}(\kappa)\Bigr]{}_{j1}\biggr\}=0\,.
\end{equation}

\subsection{generalized fluctuation-dissipation relation between the oscillators}

The fluctuation-dissipation relation of the bath field can be straightforwardly shown to hold from the definitions of the relevant Green's functions
\begin{equation}\label{E:vbnbjr}
	\overline{\mathbf{G}}_{H,0}^{(\phi)}(\kappa)=\coth\frac{\beta_{\textsc{b}}\kappa}{2}\,\operatorname{Im}\overline{\mathbf{G}}_{R,0}^{(\phi)}(\kappa)\,.
\end{equation}
The consequence of \eqref{E:rjhvjgssg}, together with the fluctuation-dissipation relation of the bath field \eqref{E:vbnbjr}, implies the fluctuation-dissipation relation of the oscillators, coupled to the bath,
\begin{equation}\label{E:vbnbjrd}
	\overline{\mathbf{G}}_{H}^{(\chi)}(\kappa)=\coth\frac{\beta_{\textsc{b}}\kappa}{2}\,\operatorname{Im}\overline{\mathbf{G}}_{R}^{(\chi)}(\kappa)\,,
\end{equation}
with the same proportionality factor $\coth\beta_{\textsc{b}}\kappa/2$, which is a function of the initial bath temperature $\beta_{\textsc{b}}^{-1}$. Thus we conclude that even the two oscillators are coupled to the common bath field with different strengths, it does not affect the equilibrium condition -- the coupling constants do not appear in the formal expression of the generalized fluctuation-dissipation relation of the oscillators. However, unequal coupling does modify the expressions of the relevant kernel functions in \eqref{E:vbnbjrd}, as can be seen in \eqref{E:ngbshkdjf}, which introduces an asymmetry in the energy flow into and out of each oscillator. In other words, for the linear systems we have investigated, the generalized fluctuation-dissipation relation is a manifestation of dynamical equilibration. This relation is of a categorical nature: it does not depend on the rate of energy exchange of each individual subsystem with its surrounding, not on the rate each subsystem approaches equilibrium, but acts as a guarantor of the global balance of power.

\section{In equilibrium, with unequal temperatures}

In the last section, we have shown the two oscillators with different couplings to a common bath can settle in equilibrium by  proving the existence of a  steady state and a generalized fluctuation-dissipation relation between the oscillators. %accounting for all three sources of power intake and output in relation to the bath and the other oscillator we have shown that each oscillator can settle down into a steady state, in equilibrium with the bath.  (strictly speaking, we did not use the energy balance,   we use the property of the poles to show the existence of the steady state in Sec.~\ref{S:gbrh})    
Here, we shall show that each oscillator with a different coupling with the bath can still be characterized by its own effective temperature which is different from the other. 
%Combining these two points we come to the stated conclusion about the zeroth law, namely, that it exists, but only in terms of the equilibrium conditions.

\subsection{Different effective temperatures of the two oscillators}

We have argued in~\cite{NEqFE} that the effective temperature \eqref{E:gbksjhd} of a single oscillator, coupled to the bath field initially at temperature $\beta_{\textsc{b}}^{-1}$, is a monotonic function of the parameter $\mathfrak{S}$
\begin{equation}
	\mathfrak{S}(t)=\langle\hat{\chi}^{2}(t)\rangle\langle\hat{p}^{2}(t)\rangle-\frac{1}{4}\langle\bigl\{\hat{\chi}(t),\hat{p}(t)\bigr\}\rangle^{2}-\frac{1}{4}\,,
\end{equation}
which is also related to the determinant of the covariance matrix for the oscillator. Hence we can also address the effective temperature of each oscillator when we have two oscillators coupled to a common bath, with the observation that $\langle\hat{\chi}_{1}^{2}\rangle=\operatorname{Tr}_{12}\{\hat{\chi}_{1}^{2}\hat{\rho}_{12}\}=\operatorname{Tr}_{1}\{\hat{\chi}_{1}^{2}\hat{\varrho}_{1}\}$, where $\hat{\rho}_{12}$ is the reduced density matrix of the combined system that contains oscillator 1 and 2, while $\hat{\varrho}_{1}=\operatorname{Tr}_{2}\hat{\rho}_{12}$ is the reduced density matrix of oscillator 1 alone.

After each oscillator is relaxed to its equilibrium state, we have
\begin{equation}
	\frac{1}{2}\langle\bigl\{\hat{\chi}_{i}(t),\hat{p}_{i}(t)\bigr\}\rangle=0\,,
\end{equation}
with $i=1$, 2. Thus the effective temperature of each oscillator will depend on the parameter $\mathfrak{S}_{i}$, which now is reduced to
\begin{equation}
	\mathfrak{S}_{i}(t)=\langle\hat{\chi}_{i}^{2}(t)\rangle\langle\hat{p}_{i}^{2}(t)\rangle-\frac{1}{4}\,.
\end{equation}
From the discussion in Sec.~\ref{S:gbrh}, it is straightforward to see that in the final equilibrium state of two oscillators, we have
\begin{align}
	\bm{\sigma}_{\chi\chi}(\infty)&=\int_{-\infty}^{\infty}\!\frac{d\kappa}{2\pi}\;\overline{\mathbf{G}}_{R}^{(\chi)}(\kappa)\cdot\mathbf{g}\cdot\overline{\mathbf{G}}_{H,0}^{(\phi)}(\kappa)\cdot\mathbf{g}^{T}\cdot\overline{\mathbf{G}}_{R}^{(\chi)\dagger}(\kappa)\,,\label{E:fksbks1}\\
	\bm{\sigma}_{pp}(\infty)&=\int_{-\infty}^{\infty}\!\frac{d\kappa}{2\pi}\;\kappa^{2}\overline{\mathbf{G}}_{R}^{(\chi)}(\kappa)\cdot\mathbf{g}\cdot\overline{\mathbf{G}}_{H,0}^{(\phi)}(\kappa)\cdot\mathbf{g}^{T}\cdot\overline{\mathbf{G}}_{R}^{(\chi)\dagger}(\kappa)\,,\label{E:fksbks2}
\end{align}
where $\bm{\sigma}_{\chi\chi}$ is defined as
\begin{equation}
	\bm{\sigma}_{\chi\chi}(t)=\frac{1}{2}\,\langle\bigl\{\bm{\Xi}(t),\,\bm{\Xi}^{T}(t)\bigr\}\rangle\,.
\end{equation}
The same applies to $\bm{\sigma}_{pp}$. Hence the diagonal elements of $\bm{\sigma}_{\chi\chi}$ and $\bm{\sigma}_{pp}$ will provide the ingredients to construct $\mathfrak{S}_{i}$ in the equilibrium state. Note that Eqs.~\eqref{E:fksbks1} and \eqref{E:fksbks2} are rather formal. In particular, the integral in \eqref{E:fksbks2} is divergent, so we need to introduce a regularization scheme  to render the integral well defined.

Let us compare the explicit  expressions of $\langle\hat{\chi}_{1}^{2}(t)\rangle$ and $\langle\hat{\chi}_{2}^{2}(t)\rangle$,
\begin{align}
	\langle\hat{\chi}_{1}^{2}(t)\rangle&=\int_{-\infty}^{\infty}\!\frac{d\kappa}{2\pi}\;\Bigl\{g_{1}^{2}\bigl[\mathbf{G}_{R}^{(\chi)}\bigr]_{11}\bigl[\mathbf{G}_{H,0}^{(\phi)}\bigr]_{11}\bigl[\mathbf{G}_{R}^{(\chi)*}\bigr]_{11}+g_{2}^{2}\bigl[\mathbf{G}_{R}^{(\chi)}\bigr]_{12}\bigl[\mathbf{G}_{H,0}^{(\phi)}\bigr]_{11}\bigl[\mathbf{G}_{R}^{(\chi)*}\bigr]_{12}\Bigr.\notag\\
	&\qquad\qquad\qquad\qquad\qquad\qquad\qquad\qquad\qquad+\Bigl.2g_{1}g_{2}\operatorname{Re}\Bigl(\bigl[\mathbf{G}_{R}^{(\chi)}\bigr]_{11}\bigl[\mathbf{G}_{H,0}^{(\phi)}\bigr]_{12}\bigl[\mathbf{G}_{R}^{(\chi)*}\bigr]_{12}\Bigr)\Bigr\}\,,\label{E:gtjfdnd}\\
	\langle\hat{\chi}_{2}^{2}(t)\rangle&=\int_{-\infty}^{\infty}\!\frac{d\kappa}{2\pi}\;\Bigl\{g_{2}^{2}\bigl[\mathbf{G}_{R}^{(\chi)}\bigr]_{22}\bigl[\mathbf{G}_{H,0}^{(\phi)}\bigr]_{22}\bigl[\mathbf{G}_{R}^{(\chi)*}\bigr]_{22}+g_{1}^{2}\bigl[\mathbf{G}_{R}^{(\chi)}\bigr]_{12}\bigl[\mathbf{G}_{H,0}^{(\phi)}\bigr]_{11}\bigl[\mathbf{G}_{R}^{(\chi)*}\bigr]_{12}\Bigr.\notag\\
	&\qquad\qquad\qquad\qquad\qquad\qquad\qquad\qquad\qquad+\Bigl.2g_{1}g_{2}\operatorname{Re}\Bigl(\bigl[\mathbf{G}_{R}^{(\chi)}\bigr]_{22}\bigl[\mathbf{G}_{H,0}^{(\phi)}\bigr]_{12}\bigl[\mathbf{G}_{R}^{(\chi)*}\bigr]_{12}\Bigr)\Bigr\}\,,\label{E:heirsk}
\end{align}
where we have suppressed the functional argument and used 
\begin{equation}
	\bigl[\mathbf{G}_{H,0}^{(\phi)}(\kappa)\bigr]_{11}=\bigl[\mathbf{G}_{H,0}^{(\phi)}(\kappa)\bigr]_{22}=\frac{\kappa}{4\pi}\,\coth\frac{\beta_{\textsc{b}}\kappa}{2}\,.
\end{equation}
Observe that similar structures also appear  in $\langle\hat{p}_{1}^{2}(t)\rangle$ and $\langle\hat{p}_{2}^{2}(t)\rangle$. In general they are different for an arbitrary choice of $g_{1}$, $g_{2}$. For example, we can make an extreme choice $g_{1}\gg g_{2}$. Thus $\mathfrak{S}_{1}$ differs from $\mathfrak{S}_{2}$, and we expect that in general both oscillators have different effective temperatures even though both of them are simultaneously coupled to the common thermal bath, and  altogether they have come to equilibrium.  

The fact that two oscillators with different couplings with the same bath can i) come to equilibrium with each other  and yet ii) each can be assigned a different effective temperature shows something interesting and fundamental.  Property i) attests to their mutual influence and ii) attests to their relative independence. This suggests that the effective temperature of each oscillator must contain some information about the other, e.g., their spatial separation. Let us look into this question. On surface, from \eqref{E:gtjfdnd} and \eqref{E:heirsk}, we may conclude that this information is only hidden in the corrections to the covariance matrix elements of a given oscillator due to the presence of the other oscillator. For example, the second and the third terms of the integrand in \eqref{E:gtjfdnd}, depend on the off-diagonal elements of the Green's function matrix, which in turn depend on the separation of the two oscillators. This spatial information will then be passed on to the effective temperature of each oscillator (its consequence will be discussed in a subsequent paper). However, in fact, the first term of the integrand in \eqref{E:gtjfdnd} also has dependence on the oscillator separation. To make this observation more explicit, we rewrite \eqref{E:ngbshkdjf} in the frequency space as
\begin{align}
	\overline{\mathbf{G}}_{R}^{(\chi)}(\kappa)=\begin{bmatrix} m(\omega_{\textsc{r}}^{2}-\kappa^{2}-i\,2\gamma_{1}\kappa) &-g_{1}g_{2}\,\overline{G}_{R,0}^{(\phi)}(\kappa;d)\\
	-g_{1}g_{2}\,\overline{G}_{R,0}^{(\phi)}(\kappa;d)&m(\omega_{\textsc{r}}^{2}-\kappa^{2}-i\,2\gamma_{2}\kappa)\end{bmatrix}^{-1}\,,
\end{align}
where $d=\lvert\mathbf{z}_{1}-\mathbf{z}_{2}\rvert$, and 
\begin{equation}
	\overline{G}_{R,0}^{(\phi)}(\kappa;d)=\int_{-\infty}^{\infty}\!d\tau\;e^{i\kappa\tau}\,G_{R,0}^{(\phi)}(\mathbf{z}_{1},t;\mathbf{z}_{2},t-\tau)=\frac{e^{i\kappa d}}{4\pi d}\,,
\end{equation}
since $G_{R,0}^{(\phi)}(\mathbf{z}_{1},t;\mathbf{z}_{2},s)$ is stationary and thus a function of $t-s$. We immediately see that
\begin{equation}
	\Bigl[\overline{G}_{R}^{(\chi)}(\kappa)\Bigr]_{11}=\overline{G}_{R}^{(\chi_{1})}(\kappa)\Bigl[1-g_{1}^{2}g_{2}^{2}\,\overline{G}_{R}^{(\chi_{1})}(\kappa)\overline{G}_{R,0}^{(\phi)}(\kappa;d)\overline{G}_{R}^{(\chi_{2})}(\kappa)\overline{G}_{R,0}^{(\phi)}(\kappa;d)\Bigr]^{-1}\,,
\end{equation}
where
\begin{align}
	\overline{G}_{R}^{(\chi_{1})}(\kappa)&=\frac{1}{m(\omega_{\textsc{r}}^{2}-\kappa^{2}-i\,2\gamma_{1}\kappa)}\,, 
\end{align}
are the retarded Green's function of the oscillator 1 alone, coupled to the thermal bath field. The same applies to $\overline{G}_{R}^{(\chi_{2})}(\kappa)$. Therefore $\bigl[\overline{\mathbf{G}}_{R}^{(\chi)}(\kappa)\bigr]_{11}$ already implicitly contains a correction due to the non-Markovian field-mediation effect, that is, information about the separation between the two oscillators.

This little demonstration highlights a fundamental principle we wish to emphasize in closing: the importance of complete self-consistency. The treatment of each component of a combined system must take into account of what happens to, and what is received from, all other parties involved, i.e., completeness and self-consistency. The fact that the FDR can play the role of a guarantor for overall balance is precisely because of this stringent requirement.

\subsection{Conclusion}

We conclude with the following summary based on the arguments we have presented  for the Gaussian quantum systems we have analyzed:

\begin{enumerate}

\item {\it Zeroth Law based on equilibrium condition}.  For quantum systems strongly coupled to a bath, a zeroth law of thermodynamics can be formulated based on the equilibrium condition, not based on equal temperature. 

\item {\it Equilibration, not thermalization}. A system may equilibrate, but not necessarily thermalize. a) Equilibrium is defined by a system  maintaining a steady state. We use multi-channel power balance to check on this condition. b) Thermal equilibrium is a stronger condition: it requires that the system obeys a Gibbsian distribution. 

\item {\it Temperature cannot be used to define or describe equilibrium state}. a) Temperature is no longer a useful signifier for, nor can it be used as a common currency in, systems in equilibrium.  b) An effective temperature can be defined for strong coupling~\cite{NEqFE}, as described in Sec.~\ref{S:etbertw}. Although it returns to the conventional temperature when the coupling is vanishingly weak, its properties are very different from those in conventional thermodynamics. 

\item {\it The effective temperature}  changes with time and varies with the coupling strength. a) In equilibrium, two subsystems may not have the same effective temperature, b) two subsystems with the same effective temperature may not be in equilibrium.

\item {\it Importance of complete self-consistency} in treating strong-coupling thermodynamics,  the mutual influences of each component of a complex system on every other component need be fully accounted for. 

\end{enumerate}

%2) in the steady configuration, two subsystems with different temperatures do not necessarily imply a steady energy flow between them, as in the NESS case, and  because now the effective temperature is time dependent. Thus they can happen to have the same value of the effective temperatures at different moments during their nonequilibrium evolutions.[Save this for later -- NESS in strong cplg]

%The zeroth law in conventional thermodynamics establishes an equivalence relation among the thermodynamic systems. It allows us to categorize the subsystems into distinct subsets, labelled by the temperature. All subsystems in a subset will, according to the zeroth law, have the same temperature. However, in the case of the finite coupling, as shown in the specific example discussed here, there is still an equivalence relations, now, in terms of the equilibrium condition, not the stronger thermal equilibrium condition. 

%The temperature is no longer a useful . In a broader context, the zero law will then be understood as an equivalence relation among systems in equilibrium, weaker than thermal equilibrium, not a relation about equal temperature.

\noindent {\bf Acknowledgments} The development of our research projects on strong coupling quantum thermodynamics is aided by recent visits of J.-T. H. to the Maryland Center for Fundamental Physics at the University of Maryland, USA, and of B. L. H. to the National Center for Theoretical Sciences and the Institute of Physics, Academia Sinica, Taiwan, ROC.

\end{document}